%% file: afshar5.tex
\documentclass{article}
\usepackage{setspace, emlines2}
\begin{document}
\large \pagestyle{plain}
\centerline{\LARGE On Visibility in the Afshar Two-Slit Experiment}\vskip 1.5cm \centerline{R. E. Kastner}
\centerline{University of Maryland, College Park} \centerline{Version of May 25, 2009} \vskip .5cm

ABSTRACT. A modified version of Young's experiment by Shahriar Afshar indirectly reveals
the presence of a fully articulated interference pattern prior to the post-selection of a particle in a ``which-slit'' basis. While this experiment does not constitute a violation of Bohr's Complementarity Principle
as claimed by Afshar, both he and many of his critics incorrectly assume that a commonly used
relationship between visibility parameter V and ``which-way'' parameter K has crucial relevance to his experiment. It is argued here that this relationship does not apply to this experimental situation and that it is wrong to make any use of it in support of claims for or against
the bearing of this experiment on Complementarity.
\newpage
\vskip 1.5cm
\input shar1.tex
\onehalfspacing

1. Background\vskip .5cm

The Afshar experiment (shown schematically in Figure 1) is a two-slit interference experiment with the addition of a lens
which focuses the beams emerging from the two slits onto detection areas
serving as ``which-slit" detectors, at plane $\sigma_2$.
At plane $\sigma_1$ in between the slits and the final detection, a thin wire grid
is placed which intercepts areas in which an interference pattern would have
minima; i.e., where the probability of particle interception at plane $\sigma_1$
is zero. Afshar uses the grid to indirectly reveal the presence of an interference
pattern at $\sigma_1$, since if no interference existed there, the intensity of detection
at the final screen would be diminished by a known amount, and this does not happen.

In typical ``which-slit" or ``which-way" experiments involving detectors of varying
efficiency placed
right behind the slits or in arms of an interferometer,
two complementary parameters V and K have traditionally been
used to describe the degree to which one can ascertain either which way the
particle went (in which case $K=1$) or one can see a fully articulated interference
pattern showing the loss of which-way information (in which case $V=1$). K and V have been shown to
obey the relationship\footnotemark[1]

$$K^2 + V^2 \leq 1 \eqno(1)$$

The relationship (1) has often been taken as an expression of Bohr's Complementarity Principle (CP),
since it displays the fact that one cannot attribute both a precise ``which-way'' property and
an interference (or ``both ways") property to a particle in such an experiment; the corresponding observables are complementary.

Since the Afshar experiment shows with high reliability that the intensity at $\sigma_2$ is not diminished when the wire grid is in place, there is indeed interference occurring at $\sigma_1$. Since interference is
clearly demonstrated and the final detection measurement is precise (i.e., with one slit blocked the particle is always detected at the detector corresponding to the open slit), Afshar concludes that both $K$ and $V$ are equal to unity, in apparent violation of the relationship (1). Indeed he uses this
apparent violation of (1) to argue that his experiment violates CP.

I have argued elsewhere (Kastner 2005) that this experiment does not show any violation
of CP since it is completely analogous to a commonplace
spin experiment in which a spin-$\frac{1}{2}$ particle is prepared in the state
$|x=+1\rangle$, confirmed to be in that state at an intermediate time through
a non-destructive measurement, and then subjected to a sharp (i.e., precise) measurement of spin along
a different (noncommuting) direction, say $z$.

Other authors agree that Afshar's experiment does not demonstrate a violation
of Complementarity. However, there seems to be some disagreement in the literature as to
exactly {\it why} the Afshar
experiment does not violate CP. Some authors argue that it has something to do
with Afshar's specific claims about fringe visibility V (e.g., Drezet, A. (2005) and Steuernagel (2005))\footnotemark[2]

The aim of this paper is
to show that arguments about V, both for and against Afshar's claims,
 are irrelevant to what is going on in this experiment.

In the next section, we discuss the result (1) and consider how it should properly be interpreted.
\vskip .5 cm

2. Visibility versus Which-Way Information: several approaches\vskip .5cm

A very simple and straightforward treatment of the tradeoff between interference and which-way information is presented
by Feynman in his (1965, pp. 3-5 and 3-6). Feynman considers a Young's experiment setup for electrons in which a light
source is positioned just downstream from the slits, so that a photon probe can be emitted whenever an electron goes
through the slits. Two appropriately placed detectors serve to detect any scattered photons, indicating which slit the electron went through. If $\phi_i$ is the amplitude for the electron to go through slit $i$ and land at screen position $x$, $a$ is
the amplitude for a photon to be scattered into the correct detector $i$, and $b$ is the amplitude for the photon to be
scattered into the incorrect detector $j$, then the probability of the electrons's detection at screen position $x$, given a photon detection at either detector, is:

$$ |a\phi_1 + b\phi_2|^2 + |a\phi_2 + b\phi_1|^2 \eqno(2)$$

(2) clearly exhibits the relationship between precise which-way detection and fringe visibility, since the former
corresponds to $b=0$ (which wipes out the interference terms) and the latter corresponds to $a=b$.

A much more general approach to this problem is presented by Englert (1996) who considers an interferometer augmented
with a which-way detection device upstream from the region of recombination of the two beams. If the which-way detector
is initially in a pure state $|d\rangle$, its state following interaction with a particle is given by $U_{\pm}|d\rangle$
where the plus/minus denotes which path is taken. Englert finds that the fringe visibility and which-way parameters $V$
and $K$ are given by\footnotemark[3]:

$$V = |\langle d|U_{-}U^\dag_{+}|d\rangle|,$$

$$K = (1 - |\langle d|U_{-}U^\dag_{+}|d\rangle|^2)^{1\over 2}  \eqno(3)$$

and thus

$$V^2 + K^2 = 1 \eqno(4)$$

Again, the precision of the which-way detection depends on the separation of the two states $U_{\pm}|d\rangle$, which is
analogous to the smallness of the photon's wavelength as it interacts with the passing electrons. A photon with a
wavelength that is large in comparison to the dimensions of the experiment will only perform a ``weak'' measurement of
slit location. Similarly if the two states $U_{\pm}|d\rangle$ overlap significantly, the which-way measurement will be
weak.

Based on Feynman's presentation, we can make the correspondence

$$V = |2ab|,$$
$$K = (1 - \vert 2ab\vert ^2)^{1\over 2},\eqno(5)$$

\noindent whence it can be seen that if $a=1, b=0$, then $V=0$, and if $a=b={1\over\sqrt 2}$, then
$V=1$.\footnotemark[4]
\vskip .5 cm

3. Analysis

The above results are straightforward implications of the quantum formalism for the experiments discussed. However,
Afshar's experiment seems to involve both a which-slit detection and fringe visibility, and he therefore claims that $V^2 + K^2 =2$. Are Feynman and Englert both wrong, or has Afshar found an interesting
loophole in their derivation? No. Their derivations only apply to the
experimental situations considered by them, in which the which-way detection occurs upstream from the region of interference, and the relationship (1) also applies only to those situations.
What this means is that Afshar is wrong in presenting his experiment as providing
 any sort of {\it interesting} violation of the relationship (1) between V and K, since the derivation
 of (1) unambiguously applies to a different experiment.

Some critics, as noted earlier, try to refute his claim to have shown a violation
of complementarity by arguing that $V \ne 1$, which tacitly
 accepts the notion that (1) is applicable and tries to argue against the
 claim that it is violated. Such arguments generally take the form of claiming that the only way to have  $V = 1$ is to have a full detection of all particles involved in the interference pattern. But these arguments miss the point, for they don't dispute that the grid successfully, if {\it indirectly}, reveals the presence of a fully articulated interference pattern
 between the slits and the final screen (since virtually no particles are blocked by the grid
 at the interference minima loci). For the argument asserting $V \ne 1$ to have any force as an upholding of eqn. (1) against Afshar's claim that it is violated,
 it needs to show that the interference pattern is distorted to the extent that an accurate slit basis
 measurement takes place at the final screen. But obviously this isn't happening---the final detection
 does nothing to eradicate or distort the interference pattern. On the contrary, the undiminished intensity of the final detection serves as indirect proof that the interference pattern remains intact in the context of a sharp post-selection measurement of a noncommuting observable.
 Thus, what Afshar shows is that you can
 post-select for a particular value of the slit basis observable and not lose interference prior
 to that post-selection. This is a different experiment than the one considered by Feynman and Englert
 and their V and K analysis does not apply.

 Interestingly, the essence of the Afshar experiment is presented by Srikanth in his (2001),
 ``Physical Reality and the Complementarity Principle." Srikanth considers a two-slit experiment in which unitarity is explicitly preserved, via an additional internal degree of freedom of the detector elements
 (which can be considered a ``vibrational'' component),
 as the amplitude contributions from each slit
 evolve toward final detection on a screen composed of those detector elements.

 The evolution of the particle + detector state from the slits to the final screen with
 initial detector state $\{|0\rangle\}$, activated detector
 spatial basis states $\{ |\phi_x\rangle \}$ and vibrational basis states $\{ |v_U\rangle , |v_L\rangle \}$ is then
 given by (where amplitudes $a_x$ and $b_x$ depend on wave number, distance, and slit of origin, and $\{|x\rangle\}$ are final particle basis states):

 $$ \frac{1}{\sqrt 2} \left(|U\rangle + |L\rangle \right) \otimes |0\rangle \rightarrow  \sum_x |x\rangle \otimes
\left[ a_x |\phi_x\rangle |v_U\rangle  + b_x |\phi_x\rangle |v_L\rangle \right] \eqno(
6)$$

 Upon detection at a particular location $x$,
 one term remains from the sum on the right-hand side of (6):

$$|x\rangle \otimes
\left[ a_x |\phi_x\rangle |v_U\rangle  +  b_x |\phi_x\rangle |v_L\rangle \right] \eqno(
7)$$

 \noindent which still, however, allows for a post-selection measurement of each detector's vibrational component
 at each detection event to obtain either $v_U$  or $v_L$ and is therefore analogous to the ``which-slit"  measurement  performed by Afshar via the lens setup. This is even more dramatic than the Afshar result because
 clearly V = 1 since a fully articulated interference pattern has been irreversibly
 recorded---not just indicated indirectly---and yet a measurement
  can be performed after the fact that seems to reveal ``which slit" the photon went through.
 However, the point is that the detector's vibrational mode {\it remains in a superposition} until that measurement
 is made, implying that each photon indeed went through both slits. As Srikanth puts it, ``...the amplitude
 contributions from both paths to the observation at [detector element] $x$ results in a superposition
 of vibrational modes. The initial superposition leaves behind a remnant superposition."
 \footnotemark[5] So, just because one can ``post-select'' by measuring the vibrational
 observable and end up with a particular corresponding
 slit eigenstate doesn't mean the particle went through that slit alone; in a very concrete
 sense, it went through both slits.

\vskip .5 cm

4. Conclusion

    The inverse relationship between V and K derived independently by both Feynman and Englert
 depends on an experimental situation in which a pre-existing superposition of slit states is ``collapsed" to some degree by a measurement of an observable whose eigenstates are components of the superposition, {\it before} the interference due to the superposition can be recorded. This collapse, characterized by an increase in K, is what causes the corresponding decrease in V.

 In the Srikanth thought
 experiment, the collapse takes place only after the interference indicating V = 1 has already been
 recorded. In the Afshar experiment, the collapse takes place after the fully articulated
 interference has been indirectly indicated to exist by the fact that the grid does not
 significantly diminish the intensity of the final detection. In cases like these, the inverse
 relationship between V and K does not apply; you can get a full interference pattern
  and than sharply post-select for a slit eigenstate. But the latter measurement doesn't give any
  physically meaningful ``which-slit" information since the particle already
  went through both slits. So thinking of K as a true ``which-way" parameter in this kind
  of post-selection is misleading,
  and it is inappropriate to argue either that
 (1) Afshar's claims about CP are wrong because his $V \ne 1$ or that (2) Afshar is right because $V = 1$ and
 $K = 1$.

\footnotetext[1] {\normalsize Cf. Feynman, 1965.}
\footnotetext[2]{\normalsize Steuernagel presents a calculation that purports to show that the visibility V in Afshar's experiment is quite low. However, this calculation seriously misrepresents both the maximum and minimum irradiances by defining them with such a low resolution that many photons counted as contributing to $I_{max}$ will also be counted as contributing to $I_{min}$. This double-counting necessarily results in little difference between $I_{max}$ and $I_{min}$ and leads to a value for V that bears little, if any, relation to the actual interference pattern.}
\footnotetext[3]{\normalsize Englert uses `D' instead of `K' for the which-way parameter.}
 \footnotetext[4]{\normalsize This formulation makes the assignment $|d\rangle = ({1\over\sqrt 2},
{1\over\sqrt 2})$ and assumes that $U_+$ and $U_-$ rotate $|d\rangle$ to $(a,b)$ and $(b,a)$, respectively, where $|a|^2
+ |b|^2 = 1$.}
\footnotetext[5]{\normalsize Srikanth ( 2001), p. 2}

\newpage
References.\vskip 1cm

\noindent Afshar, S. S. (2004), ``Sharp Complementary and Particle Behaviors in the Same Welcher Weg Experiment,''
preprint.
\newline Drezet, A. (2005), ``Complementarity and Afshar's experiment". arxiv.org: quant-ph/0508091.
 \newline Englert, B.-G.
(1996), ``Fringe Visibility and Which-Way Information: an Inequality,'' {\it Phys. Rev. Lett.} 77: 2154-7.
\newline
Feynman, R.P., R. B. Leighton, and M. Sands (1965), {\it The Feynman Lectures on Physics}, Vol. 3, Reading, MA:
Addison-Wesley.
\newline Kastner, R. (2005), ``Why the Afshar experiment does not refute complementarity". {\it Studies in History and Philosophy of Modern Physics} 36: 649-658.
\newline Srikanth, R. (2001), ``Physical Reality and the Complementarity Principle,'' arxiv.org: quant-ph/0102009.
\newline Steuernagel, O. (2005), ``Afshar's experiment does not show a violation of complementarity". arxiv.org: quant-ph/0512123.

\end{document}

%% file: shar1.tex
\unitlength 1mm 
\linethickness{0.4pt}
\ifx\plotpoint\undefined\newsavebox{\plotpoint}\fi 
\begin{picture}(138.888,115)(50,120)
\bezier{200}(106.218,235.977)(117.558,212.317)(106.218,190.977)
\multiput(106.218,235.977)(-.0327273,-.0687879){33}{\line(0,-1){.0687879}}
\multiput(105.138,233.707)(-.0331034,-.0782759){29}{\line(0,-1){.0782759}}
\multiput(104.178,231.437)(-.0326923,-.0873077){26}{\line(0,-1){.0873077}}
\multiput(103.328,229.167)(-.0331818,-.1027273){22}{\line(0,-1){.1027273}}
\multiput(102.598,226.907)(-.0326316,-.1184211){19}{\line(0,-1){.1184211}}
\multiput(101.978,224.657)(-.0333333,-.15){15}{\line(0,-1){.15}}
\multiput(101.478,222.407)(-.031667,-.1875){12}{\line(0,-1){.1875}}
\multiput(101.098,220.157)(-.03,-.248889){9}{\line(0,-1){.248889}}
\multiput(100.828,217.917)(-.03,-.446){5}{\line(0,-1){.446}}
\put(100.678,215.687){\line(0,-1){2.23}}
\put(100.638,213.457){\line(0,-1){2.23}}
\multiput(100.718,211.227)(.033333,-.37){6}{\line(0,-1){.37}}
\multiput(100.918,209.007)(.031,-.221){10}{\line(0,-1){.221}}
\multiput(101.228,206.797)(.0330769,-.17){13}{\line(0,-1){.17}}
\multiput(101.658,204.587)(.0323529,-.13){17}{\line(0,-1){.13}}
\multiput(102.208,202.377)(.033,-.11){20}{\line(0,-1){.11}}
\multiput(102.868,200.177)(.0334783,-.0952174){23}{\line(0,-1){.0952174}}
\multiput(103.638,197.987)(.0333333,-.0811111){27}{\line(0,-1){.0811111}}
\multiput(104.538,195.797)(.0336667,-.073){30}{\line(0,-1){.073}}
\multiput(105.548,193.607)(.0336667,-.0653333){30}{\line(0,-1){.0653333}}
\put(-47.09,44.147){}
\put(60.218,235.317){\line(0,-1){11.34}}
\put(60.218,217.977){\line(0,-1){11.33}}
\put(60.218,201.647){\line(0,-1){11.33}}
\put(37.443,230.953){\line(0,-1){32.33}}
\put(37.443,214.293){\vector(1,0){8}}
\put(61.218,222.977){\line(1,0){45.34}}
\multiput(106.558,222.977)(.0862765957,-.0336879433){564}{\line(1,0){.0862765957}}
\multiput(60.888,219.977)(.0906150794,-.0337301587){504}{\line(1,0){.0906150794}}
\put(106.558,202.977){\line(1,0){48.66}}
\put(60.558,202.977){\line(1,0){45.66}}
\multiput(106.218,202.977)(.0807131012,.0337313433){603}{\line(1,0){.0807131012}}
\multiput(60.558,205.317)(.0877862595,.0337022901){524}{\line(1,0){.0877862595}}
\put(106.558,222.977){\line(1,0){48.66}}
\put(96.888,197.647){\dashbox{2}(1,31)[cc]{ }}
\put(97.558,239.317){\makebox(0,0)[cc]{$\sigma_1$}}
\put(155.218,239.977){\makebox(0,0)[cc]{$\sigma_2$}}
\put(102.558,167.647){\makebox(0,0)[cc]{Figure 1. The setup for theAfshar experiment.}}
\put(57.218,220.977){\makebox(0,0)[cc]{U}}
\put(57.218,203.977){\makebox(0,0)[cc]{L}}
\put(158.888,223.317){\makebox(0,0)[cc]{L$^\prime$}}
\put(158.888,202.977){\makebox(0,0)[cc]{U$^\prime$}}
\end{picture}